\documentclass{pj}
\usepackage[pdftex]{graphicx}
\usepackage{amsmath}
\usepackage{amssymb}
\usepackage{cite}

\begin{document}
\setcounter{page}{1}
\pjheader{Vol.\ x, y--z, 2021}

\title[]
{Computational Approach of Designing Magnetfree Nonreciprocal Metamaterial}
 \footnote{\it Received date}  \footnote{\hskip-0.12in*\, Corresponding
author:~Swadesh~Poddar (poddarswadesh@gmail.com).}
\footnote{\hskip-0.12in\textsuperscript{1} Design Engineer, Qorvo, US.} 

\author{Swadesh~Poddar\textsuperscript{*, 1}, Md. Tanvir~Hasan, and Md. Ragib Shakil~Rafi}

\runningauthor{Poddar et al.}

\tocauthor{Md. Tanvir~Hasan and Md. Ragib Shakil~Rafi}

\begin{abstract}
This article aims at discussing computational approach to design magnet-free nonreciprocal metamaterial. Detailed mathematical derivation on floquet mode analysis is presented for Faraday and Kerr rotation. Non-reciprocity in the designed metasurface is achieved in the presence of biased transistor loaded in the gap of circular ring resonator. Based on the derived mathematical model, co and cross-polarized components have been extracted, which helps find Faraday and Kerr rotation and compare/contrast the reciprocal and nonreciprocal systems. 
\end{abstract}


\setlength {\abovedisplayskip} {6pt plus 3.0pt minus 4.0pt}
\setlength {\belowdisplayskip} {6pt plus 3.0pt minus 4.0pt}

\

\section{Introduction}
\label{section label}
A system is called reciprocal when signal transmission between any two ports is independent with respect to propagation direction.  Signal transmission between transmitter and receiver can be represented in terms of scattering ($S$) as $S^T = S$). In our daily life application such as antenna, passive electrical circuits, and components are reciprocal. Non-reciprocal systems, the opposite of reciprocal, exhibit diﬀerent received-transmitted field ratios when the source and observation points interchange. Non-reciprocal components such as circulators, isolators, and gyrators play a pivotal role in many microwave wireless applications, including high-power transmitters, FDD system, emerging quantum computing and readout, defense and satellite communication, to name a few \cite{Kord, caloz2018nonreciprocity}.

Since 1950, there are significant amount of researches on non-reciprocity and over time different topologies have been developed to achieve nonreciprocity in the presence of permanent magnet, using nonreciprocal component, time modulation, parametric amplifier and so on \cite{Kord, meta, Kodera1, alu_nonreciprocity, Alex}. Advances in solid state technology and significant research effort on artificially engineered material, known as metamaterial is to induce customized properties in a material that originally does not exist, have initiated a prime interest on achieving “magnet-free” non-reciprocity, where the time reversal symmetry can be broken in absence of static magnets. Over the past decade, research into metamaterials has been extended to a search for real-world applications, leading to the concept of meta-devices, defined as metamaterial-based devices that can operate in an active manner \cite{Xiao_2020}. This article will mostly aim on the computational approach of designing nonreciprocal metamaterial leading to practical realization of co-simulation and machine learning based optimization. The proposed design in this article is the extension of the transistor based design first proposed by Toshiro et al. \cite{Kodera2, ULMKodera}  and later in our previous articles a nonreciprocal metasurface was investigated with a precise electronic tunability features \cite{meta, Swadesh} as well as analytical modelling of polarization rotation in a nonreciprocal media. \cite{swadeshmathpaper}.  

The article is organized as follows. In Section II, an insight to the mechanism  of floquet mode analysis and corresponding analytical approach will be provided, in Section III, the principle of ferromagnetic resonance, and in section IV, the design approaches and process flow of our nonreciprocal devices will be described. At last in section V our discussion will be mostly focused on simulated performance, benchmarking reciprocal and nonreciprocal design, highlighting major challenges, limitation, and future works.  

\section{Floquet Mode Analysis and Mathematical Modeling}
\label{sec:formulation}

In this section, based on floquet mode analysis, a mathematical model is presented to characterize the unit cell of our designed metasurface. The Floquet port is used with planar-periodic structures. The analysis of the periodic structure is accomplished by analyzing a unit cell and master and slave boundaries are used to model a unit cell of a repeating
structure. The fields on the slave surface are constrained to be identical to those on the master surface, with phase shift. A coordinate system must be identified on the master and slave boundary to identify point-to-point correspondence and master and slave surfaces must be of identical shapes and sizes. A set of modes called floquet modes represent the fields on the port boundary. Fundamentally, Floquet modes can be assumed as plane waves where propagation direction is set by the frequency, and geometry of periodic structures. To simplify the discussion, a 3D structure has been developed and 1st 2 modes have been taken into consideration to propagate through the metamaterial unit cell. Floquet boundary at z direction has been calculated accordingly.

\vspace{0.0cm}
\begin{figure}[htbp]
\centering
    \includegraphics[height=.40\textheight]{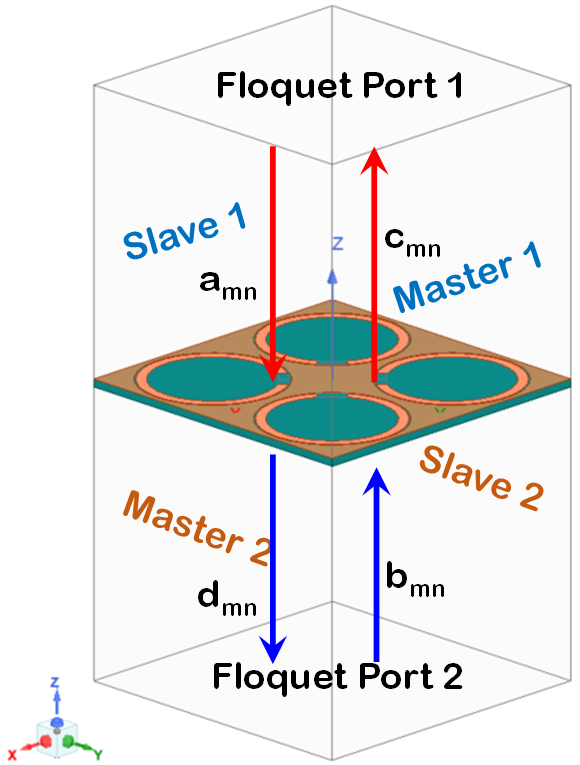}
\caption{Geometry of proposed metamaterial unit cell. Primary and secondary  master and slave boundaries form the side walls of unit cell. In both ports floquet environment scattered and incident wave is shown with arrow direction}
\label{fig:floquet_port}
\end{figure}

Figure \ref{fig:floquet_port} shows the proposed metamaterial unit cell in the floquet port environment. $a_{mn}$ and $b_{mn}$ are the incident wave from the floquet port 1 and floquet port 2, respectively. On the other hand $c_{mn}$ and $d_{mn}$ are the scattered wave.  In a normal incidence case where the 1st 2 modes, namely TE and TM is evaluated and can be expressed in S matrix. In Ansys HFSS simulation, floquet mode has been analyzed where transverse electric (TE) modes are y polarized and transverse magnetic (TM) modes are x polarized.

\begin{equation}
\left[ 
\begin{array}{c}
\mathbf{c}_{mn}^{TE} \\ 
\mathbf{c}_{mn}^{TM} \\ 
\mathbf{d}_{mn}^{TE} \\ 
\mathbf{d}_{mn}^{TM}%
\end{array}%
\right] =\left[ 
\begin{array}{cccc}
\mathbf{S}_{11}^{TE,TE} & \mathbf{S}_{11}^{TE,TM} & \mathbf{S}_{12}^{TE,TE}
& \mathbf{S}_{12}^{TE,TM} \\ 
\mathbf{S}_{11}^{TM,TE} & \mathbf{S}_{11}^{TM,TM} & \mathbf{S}_{12}^{TM,TE}
& \mathbf{S}_{12}^{TM,TM} \\ 
\mathbf{S}_{21}^{TE,TE} & \mathbf{S}_{21}^{TE,TM} & \mathbf{S}_{22}^{TE,TE}
& \mathbf{S}_{22}^{TE,TM} \\ 
\mathbf{S}_{21}^{TM,TE} & \mathbf{S}_{21}^{TM,TM} & \mathbf{S}_{22}^{TM,TE}
& \mathbf{S}_{22}^{TM,TM}%
\end{array}%
\right] \left[ 
\begin{array}{c}
\mathbf{a}_{mn}^{TE} \\ 
\mathbf{a}_{mn}^{TM} \\ 
\mathbf{b}_{mn}^{TE} \\ 
\mathbf{b}_{mn}^{TM}%
\end{array}%
\right],
\label{eq: matrix 1}
\end{equation}
where, $\mathbf{c}_{mn}^{TE}$ and $\mathbf{a}_{mn}^{TE}$ defines the scattered and incident field respectively. Besides, $\mathbf{S}_{ij}^{pol1,pol2}$, Where, i (pol1) and j (pol2) represents incident and received port (polarization) 
respectively. For example, $\mathbf{S}_{11}^{TE,TE}$ is the co polarized reflection case or in other words incident wave is TE(y) polarized from port 1 and received wave TE(y) polarized to port 1. Similarly, $\mathbf{S}_{21}^{TM,TE}$ defines a cross polarization transmission case. The above equation \ref{eq: matrix 1} can be written in a short form as below in terms of floquet modes. 

\begin{equation}
\left[ 
\begin{array}{c}
\mathbf{c}_{mn}^{TE} \\ 
\mathbf{c}_{mn}^{TM} \\ 
\mathbf{d}_{mn}^{TE} \\ 
\mathbf{d}_{mn}^{TM}%
\end{array}%
\right] =[\mathbf{S}]\cdot \left[ 
\begin{array}{c}
\mathbf{a}_{mn}^{TE} \\ 
\mathbf{a}_{mn}^{TM} \\ 
\mathbf{b}_{mn}^{TE} \\ 
\mathbf{b}_{mn}^{TM}%
\end{array}%
\right],
\end{equation}
where
\begin{equation}
\mathbf{S}=\left[ 
\begin{array}{cccc}
S(\text{FP}1:1,\text{FP}1:1) & S(\text{FP}1:1,\text{FP}1:2) & S(\text{FP}1:1,%
\text{FP}2:1) & S(\text{FP}1:1,\text{FP}2:2) \\ 
S(\text{FP}1:2,\text{FP}1:1) & S(\text{FP}1:2,\text{FP}1:2) & S(\text{FP}1:2,%
\text{FP}2:1) & S(\text{FP}1:2,\text{FP}2:2) \\ 
S(\text{FP}2:1,\text{FP}1:1) & S(\text{FP}2:1,\text{FP}1:2) & S(\text{FP}2:1,%
\text{FP}2:1) & S(\text{FP}2:1,\text{FP}2:2) \\ 
S(\text{FP}2:2,\text{FP}1:1) & S(\text{FP}2:2,\text{FP}1:2) & S(\text{FP}2:2,%
\text{FP}2:1) & S(\text{FP}2:2,\text{FP}2:2)%
\end{array}%
\right]
\label{eq: matrix2}
\end{equation}
where FP is the short form of floquet port used here in eq. \ref{eq: matrix2}. In summary, FP $p:u$, FP $q:u$ means $S_{pq}$ for $u$ polarized as incident and received wave define co polarization. Similarly, FP $p:v$, FP $q:u$ means $S_{pq}$ for incident mode as u and received wave as v polarized. In this explanation, $p$ and $q$ were used to mention port number and $v$ and $u$ to mention polarization type.


\begin{equation}
\left[ 
\begin{array}{c}
\mathbf{c}_{mn}^{TE} \\ 
\mathbf{c}_{mn}^{TM} \\ 
\mathbf{d}_{mn}^{TE} \\ 
\mathbf{d}_{mn}^{TM}%
\end{array}%
\right] =\left[ 
\begin{array}{cccc}
\mathbf{R}_{yy}^{(1)} & \mathbf{R}_{yx}^{(1)} & \mathbf{T}_{yy}^{(2)} & \mathbf{T}%
_{yx}^{(2)} \\ 
\mathbf{R}_{xy}^{(1)} & \mathbf{R}_{xx}^{(1)} & \mathbf{T}_{xy}^{(2)} & \mathbf{T}%
_{xx}^{(2)} \\ 
\mathbf{T}_{yy}^{(1)} & \mathbf{T}_{yx}^{(1)} & \mathbf{R}_{yy}^{(2)} & \mathbf{R}%
_{yx}^{(2)} \\ 
\mathbf{T}_{xy}^{(1)} & \mathbf{T}_{xx}^{(1)} & \mathbf{R}_{xy}^{(2)} & \mathbf{R}%
_{xx}^{(2)}%
\end{array}%
\right] \left[ 
\begin{array}{c}
\mathbf{a}_{mn}^{TE} \\ 
\mathbf{a}_{mn}^{TM} \\ 
\mathbf{b}_{mn}^{TE} \\ 
\mathbf{b}_{mn}^{TM}%
\end{array}%
\right]
\label{eq:matrix3}
\end{equation}

Equation \ref{eq:matrix3} has been represented in terms of reflection and transmission for the easier understanding of the floquet mode analysis. where for R/T matrix, superscript describes the received port number and subscript
describes the polarization. e.g. $T_{xx}^{1}$ equals the wave that is being
propagated from port 2 and received at port 1, for generalized
understanding, this can be mapped with $S_{12}$. In general the reflected and transmitted fields consist of both x- and y-
polarized components. Transmission field $T_{xy}$ can be defined as the
ratio of the transmitted field with polarization $x$ to the incident field with polarization $y$ Based on same concept, $R_{xy}$ can be defined as the ratio of the reflected field with polarization $x$ to the incident field with polarization $y$. 

The incident and transmitted fields are related through Jones transmission
matrix in cartesian basis as below \cite{PhysRevA.82.053811, swadeshmathpaper}

\begin{equation}
\mathbf{T}=\left[ 
\begin{array}{cc}
\mathbf{T}_{xx} & \mathbf{T}_{xy} \\ 
\mathbf{T}_{yx} & \mathbf{T}_{yy}%
\end{array}%
\right]
\end{equation}

The transmission matrix in linear basis can be mapped to transmission matrix
in circular basis as below

\begin{equation}
\mathbf{T}_{CP}=\left[ 
\begin{array}{cc}
\mathbf{T}_{\circlearrowright \circlearrowright } & \mathbf{T}%
_{\circlearrowright \circlearrowleft } \\ 
\mathbf{T}_{\circlearrowleft \circlearrowright } & \mathbf{T}%
_{\circlearrowleft \circlearrowleft }%
\end{array}%
\right] =\frac{1}{2}\left[ \mathbf{L}\right]
\end{equation}

\begin{equation}
\mathbf{T}_{CP}=\left[ 
\begin{array}{cc}
\mathbf{T}_{\circlearrowright \circlearrowright } & \mathbf{T}%
_{\circlearrowright \circlearrowleft } \\ 
\mathbf{T}_{\circlearrowleft \circlearrowright } & \mathbf{T}%
_{\circlearrowleft \circlearrowleft }%
\end{array}%
\right] =\frac{1}{2}\left[ \mathbf{L}\right] =\frac{1}{2}\left[ 
\begin{array}{cc}
T_{xx}+T_{yy}+i(T_{xy}-T_{yx}) & T_{xx}-T_{yy}-i(T_{xy}+T_{yx}) \\ 
T_{xx}-T_{yy}+i(T_{xy}\text{ }+\text{ }T_{yx}) & 
T_{xx}+T_{yy}-i(T_{xy}-T_{yx})%
\end{array}%
\right]
\end{equation}

Based on our above discussion, $\mathbf{T}$ matrix in
circular basis can be generalized as below

\begin{eqnarray}
\mathbf{T}_{\circlearrowright \circlearrowright } &=&\text{Incident RHCP to
recieved RHCP} \\
\mathbf{T}_{\circlearrowright \circlearrowleft } &=&\text{Incident LHCP to
recieved RHCP} \\
\mathbf{T}_{\circlearrowleft \circlearrowright } &=&\text{Incident RHCP to
recieved LHCP} \\
\mathbf{T}_{\circlearrowleft \circlearrowleft } &=&\text{Incident LHCP to
recieved LHCP}
\end{eqnarray}
where RHCP/LHCP stands for right/left handed circular polarization.
Once the transmission matrix in circular or linear basis is achieved, Faraday rotation and ellipticity can be calculated from there. Ellipticity, $\delta _{F}$ is a measure of the polarization state of an
electromagnetic wave and it varies from $+1$ to $-1. $In terms of RHCP and
LHCP, this can be written as%
\begin{equation}
\delta _{F}=\frac{\mid T_{\circlearrowright \circlearrowright }\mid -\mid
T_{\circlearrowleft \circlearrowleft }\mid }{\mid T_{\circlearrowright
\circlearrowright }\mid +\mid T_{\circlearrowleft \circlearrowleft }\mid }
\end{equation}
Therefore, from above equation, if $\delta _{F}= +1$, wave is RHCP and $\delta
_{F}=-1$, the wave is LHCP, whereas, $\delta _{F}= 0$ means Linear polarized wave wave.

Faraday rotation can be calculated from the equation below \cite{Lax, meta}

\begin{equation}
\theta _{F}=\frac{1}{2}\tan ^{-1}(\frac{T_{\circlearrowleft \circlearrowleft
}}{T_{\circlearrowright \circlearrowright }})
\end{equation}

Similarly, for ellipticity for a reflection based metasurface and Kerr rotation can be calculated similarly following the same steps as described above.
\begin{equation}
\delta _{K}=\frac{\mid R_{\circlearrowright \circlearrowright }\mid -\mid
R_{\circlearrowleft \circlearrowleft }\mid }{\mid R_{\circlearrowright
\circlearrowright }\mid +\mid R_{\circlearrowleft \circlearrowleft }\mid }
\end{equation}

\begin{equation}
\theta _{K}=\frac{1}{2}\tan ^{-1}(\frac{R_{\circlearrowleft \circlearrowleft
}}{R_{\circlearrowright \circlearrowright }})
\end{equation}

\section{Ferromagnetic Resonance}
\begin{figure}[htbp]
\centering
    \includegraphics[height=.15\textheight]{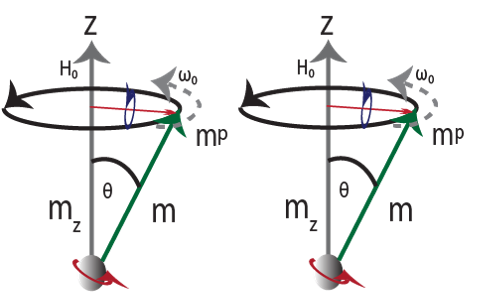}
\caption{Magnetic dipole precession, arising from electron spinning in a ferrite material about the z axis of an externally applied static magnetic bias field, $H_0$, with effective unidirectional
current loops around z axis and magnetic dipole moment $m_\rho$}.
\label{fig:E-spin}
\end{figure}

The concept related to the motion of the magnetic dipole in the presence of the magnetic field and a self-consistent RF magnetic field represents the principle idea of ferromagnetic resonance \cite{Lax}. Microwave magnetism in a ferrite material is
based on the precession of the magnetic dipole moments which arise from unpaired electron
spins along the axis of an externally applied static magnetic bias.
For resonance condition, it is necessary to apply the RF magnetic field (the microwave field) perpendicular to the dc field. The quantum mechanical phenomena can be described from Landau-Lifshitz-Gilbert equation as \cite{Lax,Landau, Swadesh}  

\begin{equation}
\frac{d\mathbf{M}}{dt} = \gamma (\mathbf{M}\times \mathbf{H})-\frac{\alpha }{\left | M \right |}\mathbf{M}\times \frac{d\mathbf{M}}{dt},
\label{eq:gyro_5}
\end{equation}
where $\alpha$ is the Gilbert damping term, $\gamma$ is the gyromagnetic ratio, $\mathbf{M}$ = $\textbf{i}_{z}$${M}_{0}$ + $\textbf{m}$$e^{j\omega t}$, and $\mathbf{H}$ = $\textbf{i}_{z}$${H}_{i}$ + $\textbf{h}$$e^{j\omega t}$, where $\textbf{i}_{z}$ is the unit vector in the z direction, $\textbf{m}$, and $\textbf{h}$ are RF quantities, and ${H}_{i}$ is the total internal dc magnetic field as shown in Fig. \ref{fig:E-spin}. The nearer the frequency of the microwave field to the natural precession frequency, the greater will be the energy absorbed by the spins \cite{Lax}. 
The longitudinal (z) component does not contribute to precession and therefore, to magnetism. Since, $\textbf{H}_{0}$ $\parallel$ $\mathbf{\widehat{z}}$, the z-component of $\textbf{M}$ produced by $\textbf{H}_{t}^{RF}$ would lead to $[M_{z}^{RF}(\mu _{0}{H}+H^{RF})](\widehat{\mathbf{z}}\times \widehat{{\mathbf{z}}})=0.$ Therefore, the only torque being produced by the transverse component at x-y plane can be stated as $(M_{t}^{RF}\mu _{0}{H})(\widehat{\mathbf{t}}\times \widehat{\mathbf{z}})\neq 0.$

\vspace{0.0cm}

The magnetization vector described above has a singularity, when $\omega$ = $\omega_{0}$ = $\gamma H_{0}$, which can be defined as the resonance condition. Resonance can be achieved by varying the operating frequency or the applied field until the precession frequency equals the microwave frequency.
Permeability tensor of ferrite material can be represented as \cite{Polder}

\begin{equation}
\overrightarrow{\mathbf{\mu}}=\mu_0
\begin{bmatrix}
\mu & -jk & 0\\ 
jk & \mu & 0\\ 
0 & 0  & 1
\end{bmatrix},
\label{eq:gyro_8}
\end{equation}
where $\mu$ = $1$ +$\frac{\omega_0 \omega_M}{\omega_0^2-\omega^2}$, $jk$ = $\frac{j\omega \omega_M}{\omega_0^2-\omega^2}$, and $\omega_M$ = $\gamma 4\pi M_0$.

\section{Nonreciprocal Device Design}
Active devices have been used in the design of magnet-free non-reciprocal metamaterials. For example, a transistor amplifier can be viewed as an isolator that amplifies
the input signal in one direction and blocks it in the reverse direction and very popular in RF microwave devices, metamaterial and antenna design \cite{swadeshring, Rabus2020}. In a series of publications between 2011 and 2019, Kodera et al. explored the unidirectional properties of transistors in the design of non-reciprocal metamaterials \cite{Kodera1,Kodera2,Kodera3, Kodera4}. A ring resonator supports two counter propagating modes with the same resonance frequency, a result of reciprocity. However, when unidirectional component such as transistor is used in the ring gap, one of the modes is blocked and the ring loaded with transistor resonates only for one type of circularly polarized waves \cite{Kord, meta}.
\vspace{0.0cm}
\begin{figure}[htbp]
\centering
    \includegraphics[height=.30\textheight]{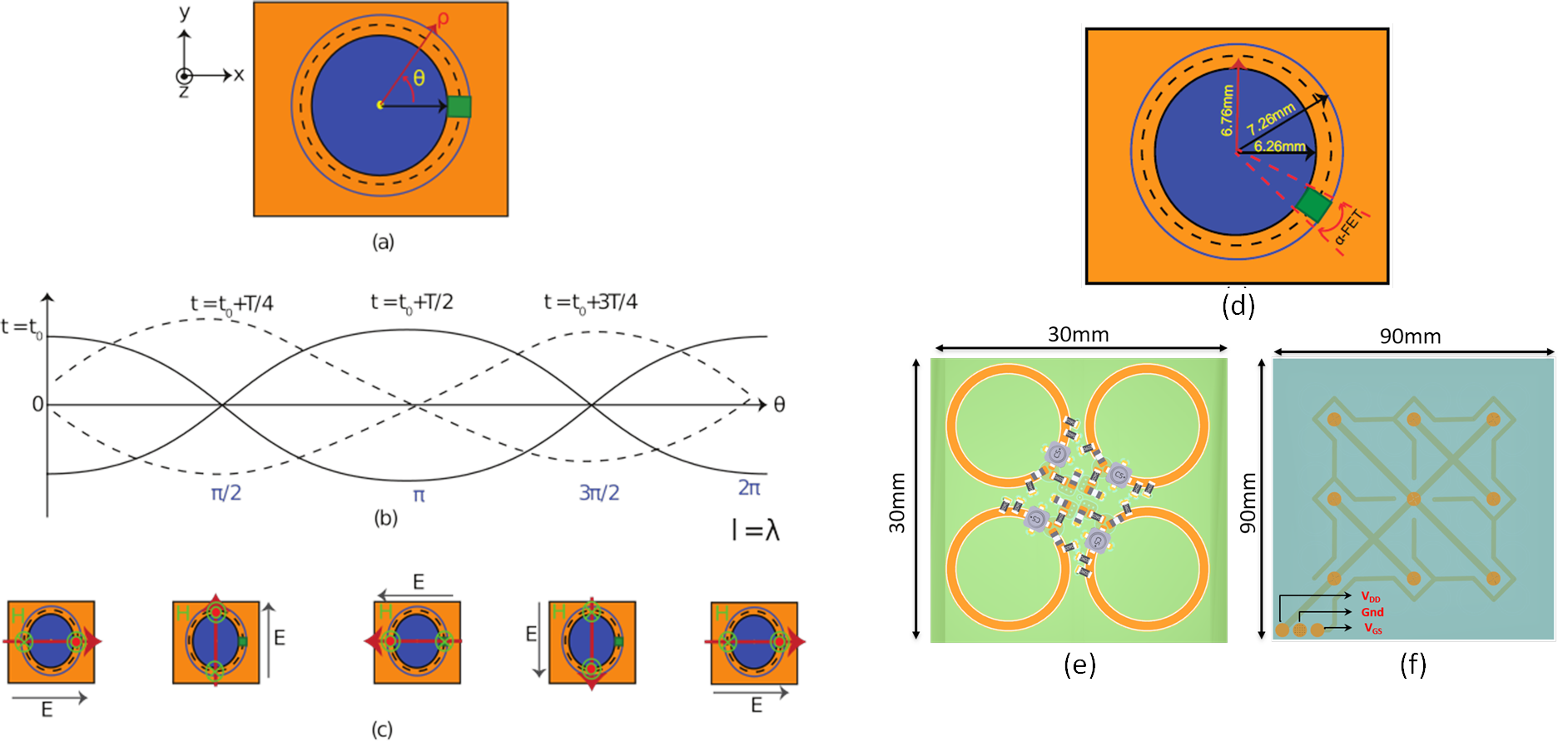}
\caption{Non-reciprocal gyrotropy. Rotating dipole moment in (a) the proposed metamaterial structure consisting of slot-ring resonator loaded with active components (transistor). (b) travelling wave along the ring at four quarter-period spaced time instants. (c) Rotating electric field in the slot-ring and the corresponding electric dipole moment due to  vectorial magnetic field pointed outward and inward. (d) Dimensions of the ring resonator (e) Unit super-cell of a 3 by 3 super-cell board consisting of 4 rings in a $90^0$-symmetric configuration (Top view). (f) Bias connection layout of the complete board (Bottom view) \cite{meta, Swadesh}}
\label{fig:rotating_vector}
\end{figure}
Figure. \ref{fig:rotating_vector} shows travelling wave resonance of a slot ring resonator loaded with transistor and passive components mimicking the electron spin precession in a biased ferromagnetic material. The electrical length ($\lambda$) of the ring resonator should be $2\pi$ due to phase matching. The wave is traveling (as opposed to standing wave) due to the presence of the unilateral (transistor) element. Based on the applied electric bias to the active component, the electric dipole moment (red arrow) rotates along the circular ring at four quarter period spaced time instants. The direction of the electric field follows the direction of the dipole moment. Therefore, non-reciprocal gyrotropy is established in the slot-ring resonator. The circular ring resonates when the electrical length is equal to an integer multiple of $2\pi$.
\begin{equation}
\beta l+\alpha _{FET}=2m\pi,
\label{eq:res_1}
\end{equation}
where $\beta$ is the propagation constant defined by $\frac{2\pi}{\lambda_g}$, $\alpha_{FET}$ is the physical distance of the ring gap as shown in Fig. \ref{fig:rotating_vector} (d), and l is
the physical length of the slotted transmission line defined as $l$ = $(2\pi - \alpha_{FET})r_{mean}$.  
In this research work, FR4 ($\varepsilon_r$ = 4.4) substrate has been used with a thickness of 0.8mm. The major focus in this research work is the Faraday rotation, an important key performance indicator to quantify nonreciprocity, which is obtained based on the transmission properties. Ground plane have been designed on geometry of the same front side so the wave can propagate through the metasurface itself. Details design modelling has been discussed in our previous work \cite{meta}.

The effective dielectric constant for a microstrip geometry is 
\begin{equation}
\varepsilon _{e}=\frac{\varepsilon _{r}+1}{2}+\frac{\varepsilon _{r}-1}{2}\left\{ 1+12\left ( \frac{H}{W} \right ) \right\}^{-\frac{1}{2}},
\label{eq:res_2}
\end{equation}
where Eq. \ref{eq:res_2} follows the condition $\frac{W}{H} \geq 1$.
and $Z_0$ can be calculated as
\begin{equation}
Z_0 = \frac{120\pi}{\sqrt{\varepsilon _{e}}\left [ \frac{W}{H}+1.393+\frac{2}{3}\ln \left ( \frac{W}{H}+1.444 \right ) \right ]}.
\label{eq:res_3}
\end{equation}
Based on the above analysis, initial parameters have been calculated considering a resonance frequency at 5.7 GHz and dielectric properties of the FR4 substrate. 

\vspace{0.0cm}
\begin{figure}[htbp]
\centering
    \includegraphics[height=.50\textheight]{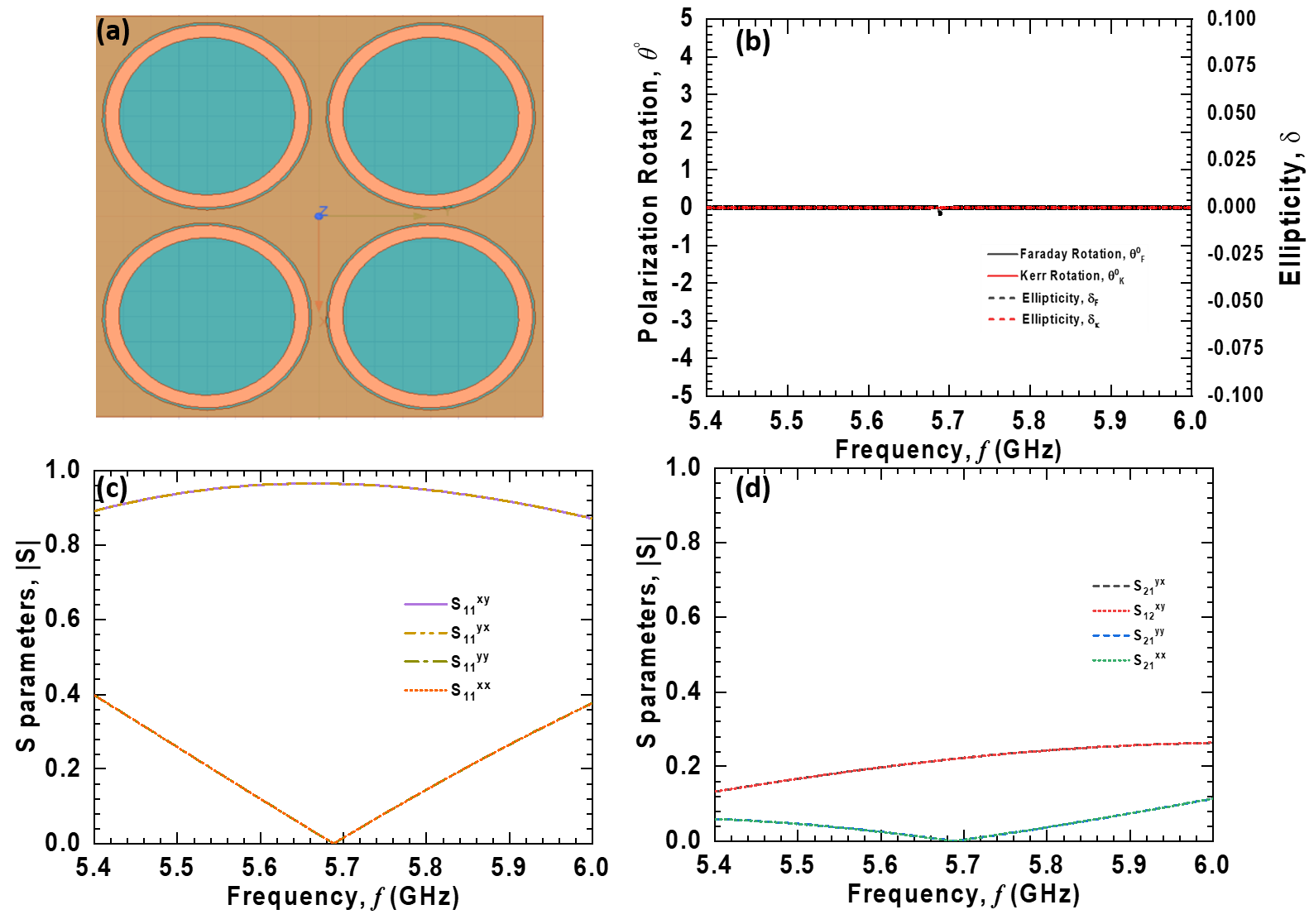}
\caption{Simulated performance of ring resonator based metamaterial. (a) Simple circular ring resonator based unit cell on proposed metasurface design. (b) The design is reciprocal, hence, no nonreciprocal properties can be observed. (c) Both co pol and (d) cross pol components exhibit the reciprocal behaviour, resonance is observed at 5.69 GHz in this design. }
\label{fig:reciprocal}
\end{figure}

Figure \ref{fig:reciprocal} (a) shows the simulated performance of a ring resonator in a reciprocal condition i.e. without any active component connected. Based on our previous analysis, nonreciprocal properties can't be achieved as shown in \ref{fig:reciprocal} (b), hence no Faraday or Kerr rotation. Propagating wave exhibits a linear polarization which can be shown from ellipticity. Figure \ref{fig:reciprocal} (c) and (d) shows reflection and transmission behavior in terms of scattering parameter. One important observation to highlight for reciprocal device is that the co and cross pol components will be in line with respect to each other. For example, we observe no difference between $S_{11}^{xy}$ and $S_{11}^{yx}$ which is cross pol reflection parameter. Co and cross pol transmission parameter follows the similar pattern.    

\vspace{0.0cm}
\begin{figure}[htbp]
\centering
    \includegraphics[height=.40\textheight]{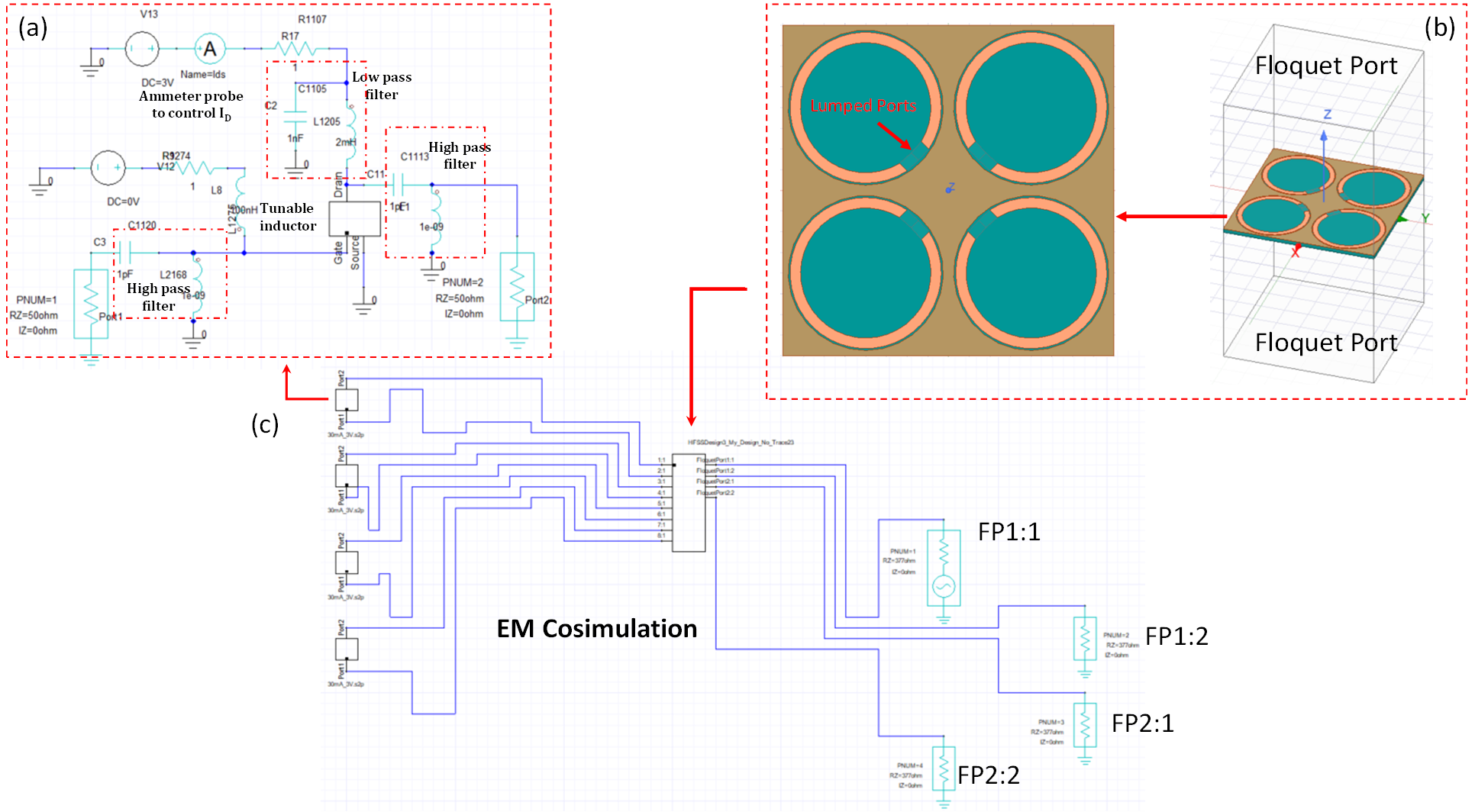}
\caption{Electromagnetic co-simulation (a) Circuit configuration that was used to achieve nonreciprocity, nonreciprocal component such as transistor was used to bridge the ring resonator gap, other component are for impedance matching and biasing the transistor in various condition (b) 3D electromagnetic model of the metamaterial using floquet mode analysis. (c) S parameter file of the 3D model and circuit model has been connected using dynamic link and to evaluate co-simulation results.} 
\label{fig:dynamink_link}
\end{figure}
Figure \ref{fig:dynamink_link} shows the setup of the co-simulation. The 3D electromagnetic structure of the metasurface have been solved in Ansys HFSS which is a passive solver. Besides, the circuit model has been solved independently in a specific bias condition. Later the S parameters are connected using the dynamic link to run the co-simulation and optimization.

\vspace{0.0cm}
\begin{figure}[htbp]
\centering
    \includegraphics[height=.25\textheight]{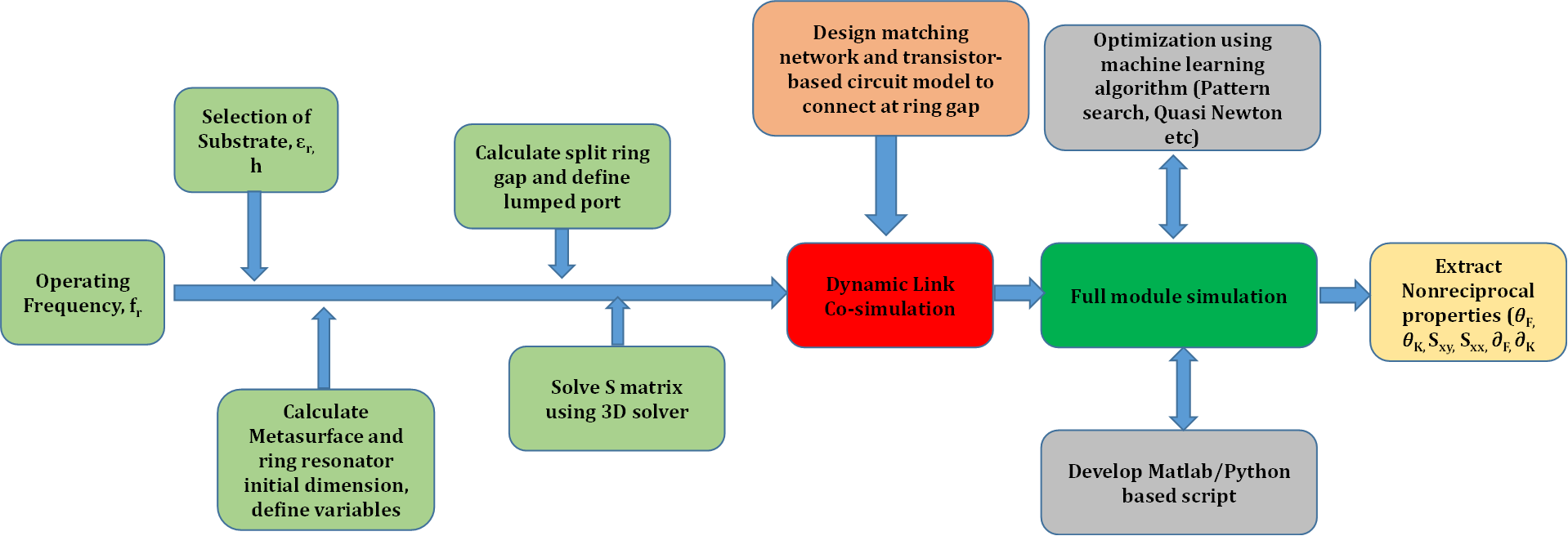}
\caption{Process flow of the computational approach of the nonreciprocal device design}
\label{fig:processflow}
\end{figure}
Figure \ref{fig:processflow} shows the process-flow of design and optimization. At first step, the frequency of operation along with preferred dielectric material is selected. After that, the initial dimension has been calculated based on microwave equation discussed earlier in the section. All dimensions are used as variables for later optimization and sensitivity analysis. Lumped ports are used in the gap and floquet based S matrix have been solved for the metasurface. Besides, in a circuit solver S parameter from our designed circuit and solved S parameter of the 3D model are cascaded for the co-simulation. Once  the initial design and result are obtained, based on the targeted goal and machine learning optimization algorithm design is optimized to achieve nonreciprocity at desired resonance. In this design, Quasi newton algorithm was used from HFSS. An in house script was developed and used during the optimization to calculate and optimize the performance at resonance on Key performance indicators such as Faraday, Kerr rotation, co-cross polarized components.   

\section{Results and Discussion}
\begin{figure}[htbp]
\centering
    \includegraphics[height=.25\textheight]{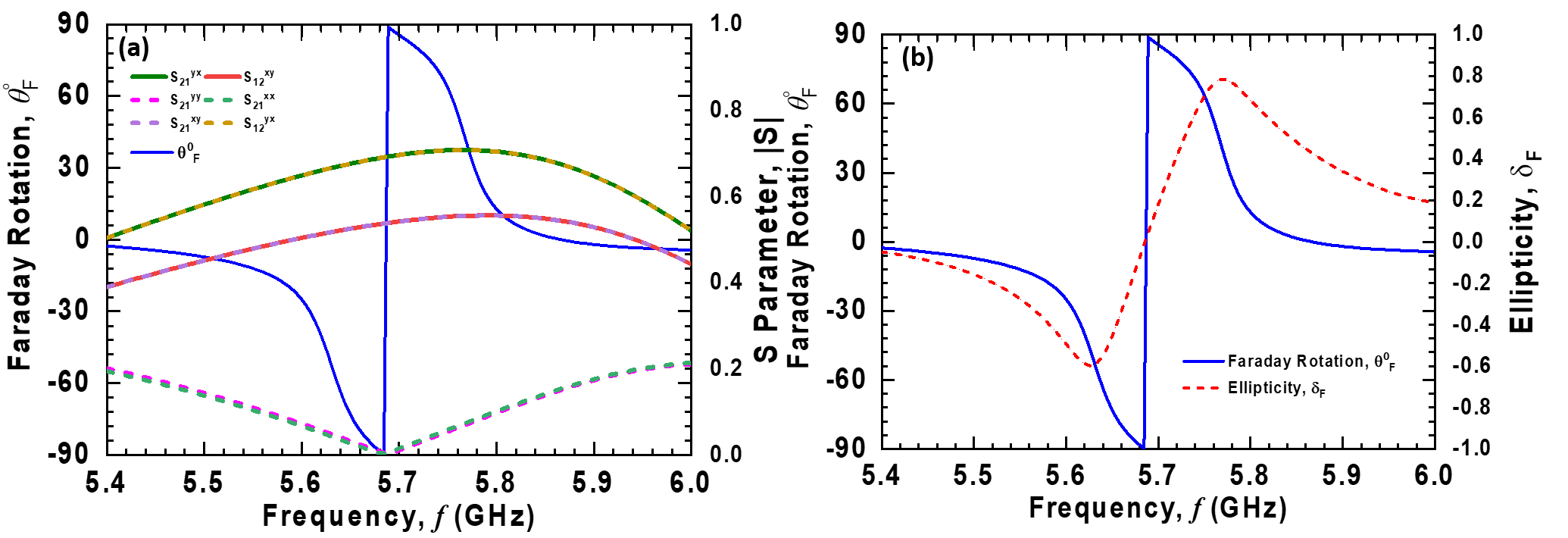}
\caption{(a) Faraday rotation and scattering parameter over frequency. co polarized components are same and at resonance those are almost 0. Cross polarized components are different and we observe maximum isolation at resonance. (b)  Faraday rotation and ellipticity over frequency.}
\label{fig:FR}
\end{figure}

\begin{figure}[htbp]
\centering
    \includegraphics[height=.25\textheight]{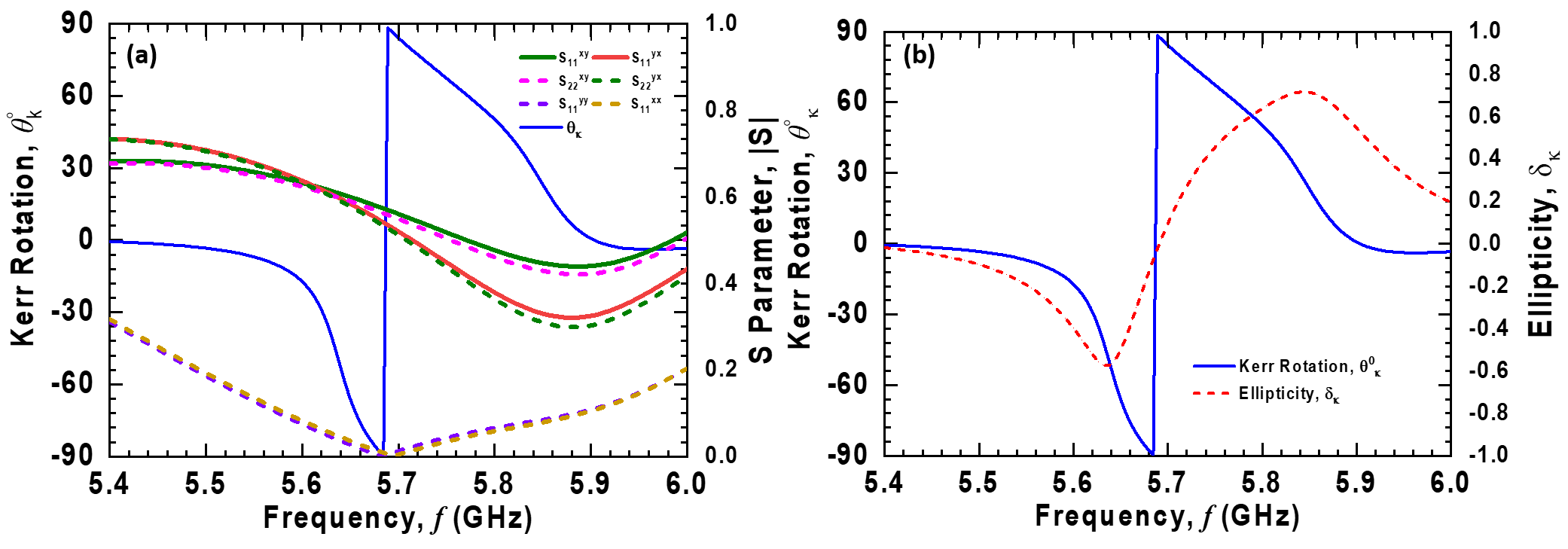}
\caption{(a) Kerr rotation and scattering parameter over frequency. co polarized components are same and at resonance those are almost 0. (b)  Faraday rotation and ellipticity over frequency.}
\label{fig:KR}
\end{figure}

In this section the simulation results of the nonreciprocal metasurface will be discussed. From antenna 1 to antenna 2, the x-polarized wave is transmitted from region 1 and y-polarized probe (antenna) pick the signal at region 2 and if the time is reversed, from antenna 2 to antenna 1, the y-polarized wave is transmitted and x-polarized probe (antenna) pick the signal at region 1. For x-polarized wave and x-polarized probing, phase of the transmission coefficients are same when the time is reversed. Therefore, likewise magnetized ferrites, the co-polarized component does not show any change in their phase when the time is reversed, however, the cross-polarized components shows $180^\circ$ phase difference at the resonance frequency so that the designed structure behaves like a magnetized ferrite where the permeability tensor has cross-components and they are out-of phase such that the non-reciprocal gyrotropic response is evident. In Fig. \ref{fig:FR} (a) the nonreciprocal behaviour is evident where we see the magnitude of the cross polarized component is different and at resonance magnitude of co polarized component is negligible. Almost $90^0$ Faraday rotation at resonance has been achieved. Similarly, from fig \ref{fig:FR} (b), at resonance $\delta_F$ = 0, i.e. the wave is linear polarized. Figure \ref{fig:KR} shows the Kerr rotation which is a reflection based measurements and can be explained similarly as we discussed for the Faraday rotation. The practical implementation and measurement procedures have been discussed in detail in our previous articles \cite{meta, TIM, swadeshmathpaper}
\vspace{0.0cm}
\vspace{0.0cm}

Design steps of a magnet-less nonreciprocal devices such as isolators, circulators, and leaky wave antennas are same however, design and optimization to achieve meaningful results are very challenging. For example, Impedance and phase matching between circular ring resonator and active/passive component at microwave frequencies are the major challenges to create artificial non-reciprocity. The excited resonance is accompanied by loss in the presence of mismatch leading to nonzero standing wave ratio (SWR). For an infinite standing wave ratio, the device does not exhibit gyrotropy, hence, non-reciprocity. Therefore, full wave 3D simulation along with optimization cost functions were used in this work and highly recommended for any non-reciprocal device design. Lastly, the number of active/passive components required to implement the artificial gyrotropy can be optimized and reduced. Characterize nonreciprocal devices are very susceptible  where precise bias condition, setup test bench and careful extraction of data is very important to take into consideration.

\section{Conclusion}
To summarize, a detailed computational approach has been presented on a fully electronic non-reciprocal metamaterial. Necessary condition to achieve nonreciprocity, analytical modelling along with key steps required for the design has been proposed, discussed and implemented. With the precision level of engineering, magnet free non-reciprocal devices can be a probable substitute of conventional ferrite material biased by permanent magnet in industrial applications such as isolator, circulator, gyrator, leaky wave antenna, and 5G applications in future. We envision that the computational approaches in this work will be very helpful in the existing research area of magnet-free non-reciprocal metamaterials.

\bibliographystyle{plain}
\bibliography{bibliography}

\begin{thebibliography}{10}

\bibitem{caloz2018nonreciprocity}
Christophe Caloz, Andrea Alù, Sergei Tretyakov, Dimitrios Sounas, Karim
  Achouri, and Zoé-Lise Deck-Léger.
\newblock Electromagnetic nonreciprocity.
\newblock {\em Physical Review Applied}, 10(4), Oct 2018.

\bibitem{Alex}
Alexander~M. Holmes, Mohsen Sabbaghi, Swadesh Poddar, Samane Pakniyat, and
  George~W. Hanson.
\newblock Experimental realization of topologically protected surface magnon
  polaritons on ceramic yig ferrites.
\newblock In {\em 2021 International Conference on Electromagnetics in Advanced
  Applications (ICEAA)}, pages 204--204, 2021.

\bibitem{ULMKodera}
T.~{Kodera} and C.~{Caloz}.
\newblock Unidirectional loop metamaterials (ulm) as magnetless artificial
  ferrimagnetic materials: Principles and applications.
\newblock {\em IEEE Antennas and Wireless Propagation Letters},
  17(11):1943--1947, 2018.

\bibitem{Kodera3}
T.~Kodera, D.~Sounas, and C.~Caloz.
\newblock Magnetless nonreciprocal metamaterial (mnm) technology: Application
  to microwave components.
\newblock {\em IEEE Transactions on Microwave Theory and Techniques},
  61:1030--1042, 2013.

\bibitem{Kodera4}
T.~{Kodera}, D.~L. {Sounas}, and C.~{Caloz}.
\newblock Switchable magnetless nonreciprocal metamaterial (mnm) and its
  application to a switchable faraday rotation metasurface.
\newblock {\em IEEE Antennas and Wireless Propagation Letters}, 11:1454--1457,
  2012.

\bibitem{Kodera2}
Toshiro {Kodera} and Christophe {Caloz}.
\newblock {Unidirectional Loop Metamaterials (ULM) as Magnetless Artificial
  Ferrimagnetic Materials: Principles and Applications}.
\newblock {\em IEEE Antennas and Wireless Propagation Letters},
  17(11):1943--1947, November 2018.

\bibitem{Kodera1}
Toshiro Kodera, Dimitrios~L. Sounas, and Christophe Caloz.
\newblock Artificial faraday rotation using a ring metamaterial structure
  without static magnetic field.
\newblock {\em Applied Physics Letters}, 99(3):031114, 2011.

\bibitem{Kord}
A.~{Kord}, D.~L. {Sounas}, and A.~{Alù}.
\newblock Microwave nonreciprocity.
\newblock {\em Proceedings of the IEEE}, 108(10):1728--1758, 2020.

\bibitem{Landau}
L.~Landau and E.~Lifshitz.
\newblock On the theory of the dispersion of magnetic permeability in
  ferromagnetic bodies.
\newblock {\em Phys. Z. Sowjetunion}, 8, 01 1992.

\bibitem{Lax}
Benjamin Lax, Kenneth Button, and H.~Hagger.
\newblock Microwave ferrites and ferrimagnetics.
\newblock {\em Physics Today - PHYS TODAY}, 16, 01 1963.

\bibitem{PhysRevA.82.053811}
Christoph Menzel, Carsten Rockstuhl, and Falk Lederer.
\newblock Advanced jones calculus for the classification of periodic
  metamaterials.
\newblock {\em Phys. Rev. A}, 82:053811, Nov 2010.

\bibitem{Swadesh}
Swadesh Poddar.
\newblock Design and analysis of fully-electronic magnet-free non-reciprocal
  metamaterial.
\newblock {\em Theses and Dissertations. 2578}, 2020.
\newblock https://dc.uwm.edu/etd/2578.

\bibitem{TIM}
Swadesh Poddar, Alexander~M. Holmes, and George~W. Hanson.
\newblock Automatic measurement technique of electromagnetic rotation in a
  nonreciprocal medium, 2022.

\bibitem{meta}
Swadesh Poddar, Alexander~M. Holmes, and George~W. Hanson.
\newblock Design and analysis of an electronically tunable magnet-free
  non-reciprocal metamaterial.
\newblock {\em IEEE Transactions on Antennas and Propagation}, pages 1--1, in
  press, 2022.

\bibitem{swadeshmathpaper}
Swadesh Poddar, Ragib~Shakil Rafi, and Md.~Tanvir Hasan.
\newblock Analytical modeling of electromagnetic rotation in nonreciprocal
  media, 2022.

\bibitem{Polder}
D.~Polder.
\newblock Viii. on the theory of ferromagnetic resonance.
\newblock {\em The London, Edinburgh, and Dublin Philosophical Magazine and
  Journal of Science}, 40(300):99--115, 1949.

\bibitem{Rabus2020}
Dominik~Gerhard Rabus and Cinzia Sada.
\newblock {\em Ring Resonators: Theory and Modeling}, pages 3--46.
\newblock Springer International Publishing, Cham, 2020.

\bibitem{swadeshring}
Sunanda Roy, M.~Abdus Samad, and Swadesh Podder.
\newblock Effect of complementary triangular split ring resonator on microstrip
  patch antenna.
\newblock In {\em 2015 2nd International Conference on Electrical Information
  and Communication Technologies (EICT)}, pages 353--358, 2015.

\bibitem{alu_nonreciprocity}
Freek Ruesink, Mohammad-Ali Miri, Andrea Alù, and Ewold Verhagen.
\newblock Nonreciprocity and magnetic-free isolation based on optomechanical
  interactions.
\newblock {\em Nature Communications}, 7(1), Nov 2016.

\bibitem{Xiao_2020}
Shuyuan Xiao, Tao Wang, Tingting Liu, Chaobiao Zhou, Xiaoyun Jiang, and Jianfa
  Zhang.
\newblock Active metamaterials and metadevices: a review.
\newblock {\em Journal of Physics D: Applied Physics}, 53(50):503002, sep 2020.

\end{thebibliography}




\end{document}